\documentclass[preprint,fleqn]{elsarticle}

\usepackage{hyperref}
\usepackage{amsmath}
\usepackage{amsfonts}
\usepackage{amssymb}

\usepackage{arydshln}
\newcommand{\su}{\uparrow}
\newcommand{\sd}{\downarrow}

\journal{Advances in Quantum Chemistry}

\begin{document}

\begin{frontmatter}

\title{Geminal theory within the seniority formalism and bi-variational principle.}

%% Group authors per affiliation:
\author[unbchem]{Stijn De Baerdemacker\corref{cor1}}
\ead{stijn.debaerdemacker@unb.ca}
\author[ugent]{Dimitri Van Neck}
\cortext[cor1]{Corresponding author}
\affiliation[unbchem]{organization={Department of Chemistry, University of New Brunswick},
                  addressline={30 Dineen Drive},
                  postcode={E3B 5A1},
                  city={Fredericton},
                  country={Canada}}
\affiliation[ugent]{organization={Center for Molecular Modeling, Ghent University},
                  addressline={Technologiepark 903},
                  postcode={9052},
                  city={Zwijnaarde},
                  country={Belgium}}

\begin{abstract}
We present an overview of the mathematical structure of geminal theory within the seniority formalism and bi-variational principle.  Named after the constellation, geminal wavefunctions provide the mean-field like representation of paired-electron wavefunctions in quantum chemistry, tying in with the Lewis picture of chemical bonding via electron pairs.  Unfortunately, despite it's mean-field product wave function description, the computational cost of computing geminal wavefunctions is dominated by the permanent overlaps with Slater determinant reference states.  We review recent approaches to reduce the factorial scaling of the permanent, and present the bi-variational principle as a consistent framework for the projected Schr\"odinger Equation and the computation of reduced density matrices. 
\end{abstract}


\end{frontmatter}

%\linenumbers

\section{A short introduction to geminals}

Geminal theory has a rich and complex history in quantum chemistry.  On the one hand, a theory built on paired electron wavefunctions touches at the very heart of chemical bonding and the Lewis picture.  On the other hand, despite their mean-field product-wavefunction interpretation, general geminal wavefunctions have been found to be perniciously difficult to compute, necessitating approximation schemes to facilitate computations of observables of interest at the potential cost of missing essential correlations.

Although originally introduced by Hurley, Lennard-Jones \& Pople \cite{hurley:1953}, the term \emph{geminals} was coined by Shull \cite{shull:1959,allen:1961} in an attempt to differentiate paired electron wavefunctions from single electron orbital wavefunctions.  Later, the term resurfaced in the context of 2-body reduced density matrix (2RDM) theory as the eigenmodes of the 2RDM \cite{coleman:1960}.  Ever since, geminals have been understood as carrying a significant component of the electron pairing structure of molecular systems.  However, on the flip side of this chemically attractive picture sits a mathematical object with the computational complexity of a \emph{permanent}, cursed with a factorial computational scaling \cite{valiant:1979,minc:1978}.  For this reason, many approximations have been suggested and proposed over the past decades in order to keep the computational complexity at bay, while retaining the chemically appealing picture of electron pairing.  A few notable examples include the antisymmetric product of strongly orthogonal geminals (APSG) \cite{surjan:1999}, the geminal power (AGP) \cite{coleman:1965}, and the more recently introduced integrable Richardson-Gaudin (RG) geminals \cite{johnson:2020} and antisymetric product of $n$-reference orbital geminals (AP$n$roG) \cite{limacher:2013}, with the $n=1$ version (AP1roG) enjoying the most attention, arguably through its simplicity and connection with pair Coupled Cluster Doubles (pCCD) \cite{stein:2014,henderson:2014b} within the projected Schr\"odinger Equation approach \cite{cullen:1996}.  All mentioned examples belong to the class of antisymmetric product of interacting geminals (APIG) \cite{silver:1969}, fitting itself within the family of seniority-zero wavefunctions \cite{racah:1943,talmi:1993}.  Within the seniority framework, a unique electron pairing structure is chosen, typically a closed shell singlet pairing within a spin-restricted single-orbital method, and electron pairs that are not compatible with this single pairing structure, such as open-shell singlets or triplets, are then considered to be broken-pair excitations.  Each unpaired electron adds one seniority unit to the wavefunction, leading to general senority-$v$ wavefunctions for $v$ unpaired electrons.  Seniority-zero wavefunctions have been proposed originally as generalized configuration interaction (CI) wavefunctions with geminal blocks \cite{allen:1962}, but have recently enjoyed more interest in an effort to systematically incorporate missing dynamical correlations into geminal approaches \cite{bytautas:2011,vanraemdonck:2015,perez:2018,kossoski:2022}. 

The present monograph provides a concise review into the mathematical structure associated with seniority-zero APIG wavefunctions within the pSE framework.  More in particular, we will focus on the application of the bi-variational principle for the computation of observables other than the energy.  A more general and comprehensive review into the potential and challenges related to AP1roG/pCCD can be found in \cite{tecmer:2022}.

This chapter is organized as follows.  A rationale for the seniority scheme and its Lie-algebraic formulation is presented in section \ref{section:seniority}, followed by an overview of variational and projective methods to compute geminals in section \ref{section:hierarchy}.  The following section presents the bi-variational principles as an internally consistent formalism to compute reduced density matrices in projective methods.  Succinct conclusions are presented in section \ref{section:conclusions}.

\section{Geminals and the seniority framework}\label{section:seniority}
\subsection{The Antisymmetrized Product of Generalized pair functions }
The most general geminal wave function is known as the Antisymmetrized Product of $N$ Generalized electron-pair functions (APG) 
\begin{equation}
|\textrm{APG}_N\rangle = \prod_{\alpha=1}^N\left[\sum_{ij=1}^{2L}G_{\alpha;ij}a_i^\dag a_j^\dag\right]|\theta\rangle\label{seniority:apg}
\end{equation}
in which the operators $a_i^\dag$ denote the creation of a single electron in one of the $2L$ orthonormal (spin-)orbitals denoted by $i$.  The creation/annihilation operators fulfill the anti-commutation relations
\begin{equation}
\{a_i^\dag, a_j^\dag\} = \{a_i, a_j\} =0,\quad \{a_i^\dag, a_j\}=\delta_{ij},\quad \forall ij=1,\dots,2L,
\end{equation}
commensurate with the antisymmetric nature of many-fermion wave functions.  $N$ is the total number of pairs. 

Geminal wave functions are ideally suited for systems with an even number of electrons, however odd numbers of electrons can be considered as well.  This can either be achieved by including an additional single-particle operator on top of the APG wavefunctions, or via an appropriate characterization of the pair-vacuum state $|\theta\rangle$.  For all practical purposes, the pair vacuum is taken to be the particle vacuum state, however it can contain any number of unpaired particles.  The tensor $G$ will be referred to as the \emph{geminal} tensor, and encodes information about how the electrons are paired across the available space of spin-orbitals.  In its most general formulation, the geminal tensor contains $N\times L^2$ free parameters, however a large portion of those is redundant, as can be seen from considering the simplest 2-electron APG state

\subsection{2-electron systems}\label{section:seniority:2electron}
For a system of 2 electrons $N=1$, the APG wave function (\ref{seniority:apg})
\begin{equation}
|\textrm{APG}_1\rangle = \sum_{ij=1}^{2L}G_{1;ij}a_i^\dag a_j^\dag|\theta\rangle\label{seniority:apg:1pair}
\end{equation}
is equivalent to the full Configuration Interaction expansion (fCI) in the basis set.  For convenience, the geminal tensor with $N=1$, can be regarded as a matrix in the orbital basis in this case. The anti-symmetric nature of the wave-function causes the symmetric part of the matrix%%
\begin{equation}
G_{1;ij} + G_{1,ji} = 0  \label{seniority:antisymmetry}
\end{equation}
to cancel exactly, leaving only the anti-symmetric part of the matrix to survive.  Interestingly, a real skew-symmetric matrix can always be brought in canonical $2\times2$ block-diagonal form by means of an orthogonal similarity transformation \cite{wimmer:2012,golub:1996}
\begin{equation}
O^TG_{1}O = \left(\begin{array}{ccccc} 0 & \lambda_1 & \dots  & &  \\
                                        -\lambda_1 & 0 &  &  &  \\
                                        \vdots &  & 0 & \lambda_2 &  \\
                                         & & -\lambda_2 & 0 & \\
                                         & & & & \ddots \end{array}\right),\label{seniority:1pair:transfo}
\end{equation} 
with more information on the algorithm how to obtain the matrix $O$ taken from \cite{wimmer:2012}.  Applying the transformation $O$ to the fermion operators
\begin{equation}
c^\dag_k = \sum_{i=1}^{2L}O^T_{ki}a_i^\dag,
\end{equation}
we can rewrite the 1-pair APG as 
\begin{equation}
|\textrm{APG}_1\rangle = \sum_{i=1}^{L}\lambda_ic_{2i-1}^\dag c_{2i}^\dag|\theta\rangle
\end{equation}
As a result, a 2-electron wavefunction can always be interpreted as a linear combination of independent pair creation operators
\begin{equation}
S_i^\dag = c_{2i-1}^\dag c_{2i}^\dag.
\end{equation}
How the electrons will pair together is dictated by the geminal matrix $G_1$.  As such, the pairing scheme can be completely general, however for spin-symmetric Hamiltonians, the pairing scheme will always be such that spin-up and spin-down electrons in the same spatial orbitals will pair together.  Therefore the spin-orbitals $2i-1$ and $2i$ will share the same spatial orbital, denoted by $i$, and only differ in spin
\begin{equation}
S_i^\dag = a_{i\su}^\dag a_{i\sd}^\dag.\label{seniority:paircreation}
\end{equation}
To fix ideas in the remainder of the chapter, the pairing operator will assume a pairing scheme in the same spatial orbital, which would be consistent with a closed-shell interpretation in the restricted Hartree-Fock sense, although general pairing schemes are equally possible in the open-shell unrestricted or even generalized Hartree-Fock sense \cite{johnson:2017}. 
\subsection{The seniority quantum number}
The definition of the independent pair operators (\ref{seniority:paircreation}) leads naturally to the introduction of the seniority quantum number via the following representation \cite{racah:1943,talmi:1993} 
\begin{equation}
S_i^\dag = a_{i\su}^\dag a_{i\sd}^\dag,\quad S_i=(S_i^\dag)^\dag = a_{i\sd}a_{i\su},\quad S_i^0 = \tfrac{1}{2}(a_{i\su}^\dag a_{i\su}+a_{i\sd}^\dag a_{i\sd}-1),\label{seniority:seniority:operators}
\end{equation}
of the $su(2)$ \emph{quasi-spin} Lie-algebra
\begin{equation}
[S_i^0,S_i^\dag]=S_{i}^\dag,\quad [S_i^0,S_j]=-S_{i},\quad [S_i^\dag,S_i] = 2S_i^0.\label{seniority:seniority:quasispin}
\end{equation}
Each spatial orbital $i$ defines an independent $su(2)_{i|S}$ quasi-spin algebra, as all pairing algebras commute mutually 
\begin{equation}
[S_i^\dag,S_j^\dag]=[S_i^0,S_j^\dag]=[S_i^\dag,S_i]=[S_i^0,S_j^0]=0,\quad\forall i\neq j.
\end{equation}
Although intuitively evident, these communation relations mark a subtle but important difference between the single-particle creation/annihilation operators $\{a_i^\dag,a_i\}$ and electron-pair creation/annihilation operators $\{S_i^\dag,S_i\}$, as the former don't commute and the latter don't anticommute between different (spin)-orbitals 
\begin{equation}
[a_i^\dag,a_j^\dag]=2a_i^\dag a_j^\dag\neq0,\quad\&\quad\{S_i^\dag,S_j^\dag\}= 2S_i^\dag S_j^\dag \neq0,\quad\forall i\neq j.
\end{equation}
For this reason, electron pairs are often referred to as \emph{hard-core} bosons.  The hard-core aspect stems from the impossibility to stack more than two pairs into one spatial orbital despite their bosonic symmetric nature, due the the nilpotency of the creation operator
\begin{equation}
S_i^\dag S_i^\dag = 0.
\end{equation}
One can now identify an $su(2)_{i|S}$ algebra with each individual spin-orbital $i$, and as they are all independent from each other, the full Lie-algebra is a kronecker sum of the individual algebras
\begin{equation}
\bigoplus_{i=1}^L su(2)_{i|S} = su(2)_{1|S}\oplus su(2)_{2|S}\oplus\dots\oplus su(2)_{L|S}.
\end{equation}

Identifying the representations of each individual $su(2)_{i|S}$ elucidates the partitioning of the Hilbert space in terms of pair creation operators.  The representations are characterized by the quantum numbers of the $su(2)_{i|S}$ Casimir and projection operators 
\begin{align}
\mathcal{C}_{2i|S}|s_i,m_i\rangle&=s_i(s_i+1)|s_i,m_i\rangle\\
S_i^0|s_i,m_i\rangle&=m_i|s_i,m_i\rangle
\end{align}
with the projection operator defined by eqs.\ (\ref{seniority:seniority:operators}), and the Casimir operator given by
\begin{equation}
\mathcal{C}_{2i|S}=\tfrac{1}{2}[S_i^\dag S_i + S_i S_i^\dag] + (S_i^0)^2 = \tfrac{3}{4}[1-(n_{i\su}-n_{i\sd})^2]\label{seniority:seniority:casimir}
\end{equation}
with the number operators defined as
\begin{equation}
n_{i\su}=a_{i\su}^\dag a_{i\su},\quad n_{i\sd}=a_{i\sd}^\dag a_{i\sd}
\end{equation}
The number operators in the Casimir operator (\ref{seniority:seniority:casimir}) allow for the definition of the \emph{seniority} operator 
\begin{equation}
v_i=(n_{i\uparrow}-n_{i\downarrow})^2,
\end{equation}
which counts the number of electrons that are \emph{not} paired in spin-orbital $i$.  The seniority quantum number $v_i=0$ if the spin orbital is either empty or doubly occupied, and $v_i=1$ if only one of the spin-up or spin-down orbitals is occupied.  This allows one to separate the four possible occupation states of a spin-orbital $i$ into two different $su(2)_{i|S}$ representations, with quantum numbers
\begin{equation}
|s_i,m_i\rangle=|\tfrac{1}{2}(1-v_i),\tfrac{1}{2}(n_{i\su}+n_{i\sd}-1)\rangle.
\end{equation}
More specifically, we have the $v_i=0$ representation states, given by the particle vacuum and fully paired state
\begin{equation}
|\theta\rangle = |\tfrac{1}{2},-\tfrac{1}{2}\rangle,\quad S_i^\dag|\theta\rangle =S_i^\dag|\tfrac{1}{2},-\tfrac{1}{2}\rangle= |\tfrac{1}{2},\tfrac{1}{2}\rangle,
\end{equation}
and the twofold degenerate single-particle states with $v_i=1$ leading to
\begin{equation}
a_{i\su}^\dag|\theta\rangle = |0_{\su},0\rangle,\quad a_{i\sd}^\dag|\theta\rangle = |0_{\sd},0\rangle,
\end{equation}
in which the notation $s_i=0_{\su}/0_{\sd}$ is employed to differentiate between the two degenerate trivial representations.  Both states can indeed be interpreted as trivial singlet $su(2)$ representations as the action of the pair creation/annihilation operators both annihilate the state
\begin{equation}
S_i^\dag|0,0\rangle = S_i|0,0\rangle=0.
\end{equation}
Finally, since all  $su(2)_{i|S}$ quasi-spin algebras are independent, one can categorize the full Hilbert space by means of its independent representations
\begin{equation}
\bigotimes_{i=1}^L|s_i,m_i\rangle= \bigotimes_{i=1}^L|\tfrac{1}{2}(1-v_i),\tfrac{1}{2}(n_{i\su}+n_{i\sd}-1)\rangle
\end{equation}
each one categorized by means of their own seniority quantum number $v_i$.  For instance, a fully paired state will be represented by one of 
\begin{equation}
\bigotimes_{i=1}^L|\tfrac{1}{2},m_i\rangle= \bigotimes_{i=1}^L|\tfrac{1}{2},\tfrac{1}{2}(n_{i\su}+n_{i\sd}-1)\rangle\label{seniority:seniority:doci}
\end{equation}
in which $n_{i\su}+n_{i\sd}$ is either 0 (empty) or 2 (fully paired).  As each spin-orbital has $v_i=0$, these states are also referred to as general seniority-zero states.  The full space of all seniority-zero states is also known as the Doubly-Occupied Configuration Interaction (DOCI) space \cite{weinhold:1967,bytautas:2011}, as only zero- or doubly occupied spin-orbitals span the space.  
\subsection{Dual spin quantum numbers}
An interesting property of the seniority-zero space is that it also carries a total spin-zero representation of the wavefunction.  This is directly related to the dual nature of the $su(2)_{i|S}$ quasi-spin algebra (\ref{seniority:seniority:operators}) \& (\ref{seniority:seniority:quasispin}) with the $su(2)_{i|J}$ spin algebra realized by
\begin{equation}
J_i^\dag = a_{i\su}^\dag a_{i\sd},\quad J_i=(J_i^\dag)^\dag = a_{i\sd}^\dag a_{i\su},\quad J_i^0 = \tfrac{1}{2}(n_{i\su}-n_{i\sd}),
\end{equation}
with commutation relations 
\begin{equation}
[J_i^0, J_i^\dag]=J_i^\dag,\quad[J_i^0, J_i]=-J_i,\quad[J_i^\dag,J_i]=2J_i^0,
\end{equation}
and Casimir operator
\begin{equation}
\mathcal{C}_{2i|J}= \tfrac{3}{4}(n_{i\su}-n_{i\sd})^2.\label{seniority:spin:casimir}
\end{equation}
Similar as the quasi-spin algebras, all the $su(2)_{i|J}$ spin algebras are mutually independent
\begin{equation}
[J_i^\dag,J_j^\dag]=[J_i^0,J_j^\dag]=[J_i^\dag,J_i]=[J_i^0,J_j^0]=0,\quad\forall i\neq j,
\end{equation}
and moreover, all quasi-spin $su(2)_{i|J}$ and $su(2)_{i|J}$ spin algebras are mutually independent as well
\begin{equation}
[J_i^\dag,S_j^\dag]=[J_i^0,S_j^\dag]=[J_i,S_j^\dag]=[J_i^\dag,S_j^0]=[J_i^0,S_j^0]=0,\quad\forall i, j,
\end{equation}
\emph{including} for $i=j$.  This leads to the identification of a larger $o(4)_i$ algebra, isomorphic to the direct sum of both $su(2)_i$ algebras
\begin{equation}
o(4)_i \cong su(2)_{i|S}\oplus su(2)_{i|J},  
\end{equation} 
which in itself is a subalgebra of an $o(5)_i\equiv sp(2,\mathbb{R})_i$ algebra
\begin{equation}
o(5)_i\supset o(4)_i \cong su(2)_{i|S}\oplus su(2)_{i|J},  
\end{equation} 
obtained by incorporating the single-particle creation/annihilation operators in the algebra \cite{iachello:2006,debaerdemacker:2009b}
\begin{equation}
T_{i,\frac{1}{2}\frac{1}{2}}=\tfrac{1}{2}a^\dag_{i\su},\quad T_{i,-\frac{1}{2}\frac{1}{2}}=\tfrac{1}{2}a_{i\sd},\quad T_{i,\frac{1}{2}-\frac{1}{2}}=\tfrac{1}{2}a^\dag_{i\sd},\quad T_{i,-\frac{1}{2}-\frac{1}{2}}=\tfrac{1}{2}a_{i\su}\label{seniority:dual:toperators}
\end{equation}
As the two $su(2)_i$ algebras commute, the four possible states of a single spin-orbital can be unique labelled by both sets of quantum numbers
\begin{align}
|\theta\rangle = |\tfrac{1}{2},-\tfrac{1}{2};0,0\rangle, &\quad a_{i\su}^\dag|\theta\rangle = |0,0;\tfrac{1}{2},+\tfrac{1}{2}\rangle\label{seniority:dual:4states1}  \\
a_{i\su}^\dag a_{i\sd}^\dag|\theta\rangle = |\tfrac{1}{2},+\tfrac{1}{2};0,0\rangle, &\quad a_{i\sd}^\dag|\theta\rangle = |0,0;\tfrac{1}{2},-\tfrac{1}{2}\rangle,\label{seniority:dual:4states2} 
\end{align}
in which the first pair of quantum numbers denote the $su(2)_{i|S}$ quasi-spin representation, and the second pair the $su(2)_{i|J}$ spin representation.  The duality between the two $su(2)_i$representations is beautifully illustrated from the relation between the Casimir operators (\ref{seniority:seniority:casimir}) \& (\ref{seniority:spin:casimir})
\begin{equation}
\mathcal{C}_{2i|S}+\mathcal{C}_{2i|J}=\tfrac{3}{4},
\end{equation}
pointing out that for each state, either the quasi-spin or the spin representation must be the doublet or singlet, but never both at the same time.  As a corollary, the seniority-zero states (\ref{seniority:seniority:doci}) must be spin-zero states as well. 

\begin{figure}[!htb]
\begin{center}
\includegraphics{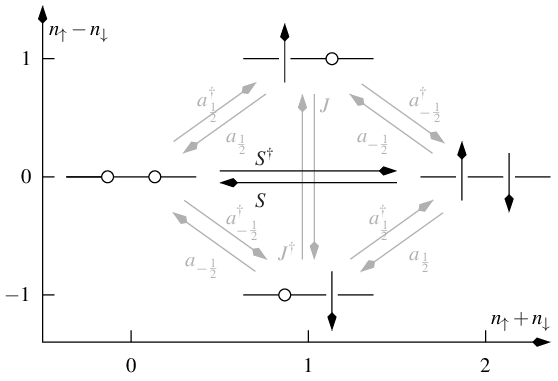}
\caption{The action of the quasi-spin (black), spin (gray) and single-particle creation/annihilation operators (gray) on the four allowed states of a individual spin-orbital.}\label{fig:seniority}
\end{center}
\end{figure}

From their notation, it is apparent that the $T$-operators in eq.\ (\ref{seniority:dual:toperators}) have a bi-spinor tensorial character in both the quasi-spin and spin space \cite{debaerdemacker:2009b,debaerdemacker:2007b}.  This means that the action of one of the $T$-operators, or equivalently the single particle creation/annihilation operators, will not only alter the spin quantum number with a unit of $\pm\frac{1}{2}$, but also the quasi-spin.  From a chemical point of view, this means that creating/annihilating single particles will alter the seniority quantum number of a state with $\pm1$, depending on the seniority of the reference state and the particular $T$-operator.  For instance, the action of a single-particle creation operator on the seniority-zero vacuum state $|\theta\rangle$ will give rise to a seniority-one (or quasispin $0$) state in both cases.  This property will prove important when extracting DOCI Hamiltonians.  The action of the $o(5)_i$ algebra generators on the four states are shown in Figure \ref{fig:seniority}
\subsection{Doubly-Occupied Configuration Interaction}
The DOCI space restricts the full Configuration Interaction (fCI) space to only the seniority-zero representations.  The appeal of DOCI space is not only that it narrows the space to only the relevant paired-electron representations, it also means a significant reduction of the fCI space.  This is most apparent when restricting the fCI space to the $S_z=0$ subspace of states which holds as many spin-up as spin-down particles.  In this case, the dimension of the Hilbert subspace is 
\begin{equation}
\dim\mathcal{H}_{\textrm{fCI}|S_z=0}=\binom{L}{N}_{\su}\times \binom{L}{N}_{\sd},
\end{equation}   
which is the square of the DOCI space in which only $N$ pairs need to be distributed over $L$ spatial orbitals
\begin{equation}
\dim\mathcal{H}_{\textrm{DOCI}}=\binom{L}{N}=\sqrt{\dim\mathcal{H}_{\textrm{fCI}|S_z=0}}.
\end{equation} 
Although DOCI is still plagued by a combinatorial scaling of its Hilbert space, it does give us a significant reduction with respect to fCI, allowing for the diagonalization of systems that are beyond the capacity of the latter.  

The DOCI Hamiltonian can be derived from the fCI Hamiltonian by restricting it to the seniority-zero space only.  The fCI Hamiltonian is given by
\begin{equation}
H = \sum_{ij=1}^{2L} \langle i|T+V_1|j\rangle a_i^\dag a_j +\frac{1}{2}\sum_{ijkl=1}^{2L}\langle ij|V_2|kl\rangle a_i^\dag a_j^\dag a_l a_k,\label{seniority:doci:fcihamiltonian}
\end{equation}
with the 1- and 2-body matrix elements are given by
\begin{align}
\langle i|T+V_1|j\rangle &=\int \phi_i^\ast(\vec{x})[T+V_1(\vec{x})]\phi_j(\vec{x})d\vec{x},\\
\langle ij|V_2|kl\rangle &= \int \phi^\ast_i(\vec{x}_1)\phi^\ast_j(\vec{x}_2)V_2(\vec{x}_1,\vec{x}_2)\phi_k(\vec{x}_1)\phi_l(\vec{x}_2)d\vec{x}_1d\vec{x}_2,
\end{align}
in which the coordinates $\vec{x}$ denote a full set of 1-particle variables, typically the spatial and spin coordinates $\vec{x}=\{\vec{r},\sigma\}$.  For reasons of generality, the 1-body and 2-body interactions $V_1$ and $V_2$ have been left unassumed for now, as DOCI Hamiltonians are applicable to a variety of physical systems, including condensed matter \cite{bardeen:1957,phillips:2003} and nuclear physics \cite{bohr:1958,dean:2003}.  Focusing on quantum chemistry, the Hamiltonian (\ref{seniority:doci:fcihamiltonian}) is the non-relativistic and spin-symmetric Coulomb Hamiltonian, so the pairing will be assumed to happen within the spin-up and spin-down partners of the spatial orbitals.  Filtering out the terms that are seniority preserving in the fCI Hamiltonian, gives rise to the DOCI Hamiltonian
\begin{equation}
H_{\textrm{DOCI}} = \sum_{i=1}^L \varepsilon_i n_i + \sum_{i,k=1}^L V_{ik} S_i^\dag S_k + \sum_{i,k=1}^L W_{ik}n_i, n_k\label{seniority:doci:hamiltonian}
\end{equation}
in which the pairing operators are defined as before (\ref{seniority:seniority:operators}) and the number operators $n_i = n_{i\su}+n_{i\sd}$, and the interaction matrix elements are given by
\begin{align}
\varepsilon_i &= \langle i |T+V_1|i\rangle,\\
V_{ik}&=\langle ii||kk\rangle, \\
W_{ik}&=\tfrac{1}{4}[2\langle ik||ik\rangle - \langle ik||ki\rangle](1-\delta_{ik}),
\end{align}
with $\langle ij||kl\rangle$ the 2-body Coulomb repulsion integrals in physical notation
\begin{equation}
\langle ij||kl\rangle = \int \phi^\ast_i(\vec{r}_1)\phi^\ast_j(\vec{r}_2)\frac{1}{|\vec{r}_1-\vec{r}_2|}\phi_k(\vec{r}_1)\phi_l(\vec{r}_2)d\vec{r}_1d\vec{r}_2.
\end{equation}
The diagonal shape of the 1-body operator is fairly straightforward to understand as all non-diagonal 1-body terms will give rise to seniority-altering terms.  A similar argument holds for the reduction of the traditional 4-point integrals to the 2-point interaction matrices $V$ and $W$, in which the $(1-\delta_{ik})$ arises to avoid double counting. 

It should be clear that the seniority-zero filtering of the fCI Hamiltonian is dependent on the spatial orbitals.  Standard DOCI computations are understood to be framed in the basis of the canonical Hartree-Fock orbitals.  However, this basis will not necessarily yield the best ground-state energy with respect to the exact fCI ground-state energy. This is commonly referred to as the orbital-optimization (OO) process associated with DOCI, which aims to minimize the DOCI ground-state energy over all possible spatial orbital sets.  Around molecular equilibrium, the optimized orbitals will be very alike the Hartree-Fock orbitals, however molecular systems in bond-breaking regimes and beyond will prefer localized sets of orbitals \cite{limacher:2013}.  Several approaches to deal with orbital optimization in seniority-zero spaces have been proposed in the literature, as they are an important aspect in geminal theory \cite{boguslawski:2014b,boguslawski:2014c,poelmans:2015,alcoba:2014a,alcoba:2014b}.
\subsection{The Antisymmetrised Product of Interacting Geminals}

The DOCI Hamiltonian still confronts us with a diagonalization procedure that scales combinatorially in the system size.  For this reason, it is important to look for compact wavefunction approximations.  The APG wavefunction (\ref{seniority:apg}) is very much appealing, as it has a pronounced mean-field flavor to it, in line with the Slater determinant mean-field representation in Hartree-Fock theory.  However, there are some major caveats.  Unless all $G_{\alpha}$ matrices ($\alpha=1,\dots,N$) in the APG commute mutually, each electron pair has a different pairing scheme in the spatial orbitals, given by the transformation (\ref{seniority:1pair:transfo}).  Therefore, the general APG wavefunction consists of a combinatorial number of seniority-breaking terms, voiding the benefits that would have come from a clean pairing theory \cite{mcweeny:1963}. 

Therefore, it makes sense to stay within a single pairing framework, and reduce the general APG wavefunction to an Antisymmetrized Product of \emph{Interacting} Geminals (APIG), originally proposed by Silver \cite{silver:1969,silver:1970}
\begin{equation}
|\textrm{APIG}_N\rangle = \prod_{\alpha=1}^N\left[\sum_{i=1}^{L}G_{\alpha i}a_{2i-1}^\dag a_{2i}^\dag\right]|\theta\rangle=\prod_{\alpha=1}^N\left[\sum_{i=1}^{L}G_{\alpha i}S_i^\dag\right]|\theta\rangle\label{seniority:apig}
\end{equation}
in which all electron pairs are expressed in the same pairing framework.  As a result, each separate generalized pair creation operator is seniority-conserving, giving rise to seniority-zero state. In the remainder of this manuscript, we will refer to the $N\times L$ matrix $G$ as the geminal matrix, rather than the geminal tensor
\begin{equation}
G=\left(\begin{array}{cccc:ccc} G_{11} & G_{12} & \dots & G_{1N} & G_{1,N+1} & \dots & G_{1L}\\ G_{21} & G_{22} & \dots & G_{2N} & G_{2,N+1} & \dots & G_{2L}\\ \vdots & \vdots & & \vdots & \vdots & & \vdots \\ G_{N1} & G_{N2} & \dots & G_{NN} & G_{N,N+1} & \dots & G_{NL}\end{array}\right).
\end{equation}
The vertical dashed line after the $N$-th column marks the square part of the matrix and serves as a reminder that the matrix is rectangular in general.
\section{A hierarchy of geminals}\label{section:hierarchy}
\subsection{permanents}
A major appeal of single-particle Hartree-Fock mean field theory is its computational scaling and tractability within a variational formalism, all related to the salient algebraic properties of determinants.  The central role of the determinant in Hartree-Fock theory is apparent from the overlap between two different Slater determinant states, which reduces to the \emph{determinant} of the coefficient matrix of one Slater determinant expressed in the basis of the other.  The determinant of a matrix remains invariant under unitary transformations, so it is always within computational reach at a polynomially scaling linear algebra manipulation, such as diagonalisation for Hartree-Fock. 

A similar property holds for APIG states (\ref{seniority:apig}), however with a subtle difference which holds catastrophic consequence for tractability.  To appreciate this, we only need to consider the overlap of the APIG states with respect to the seniority-zero Slater determinants
\begin{equation}
|[N_1,N_2,\dots,N_L]\rangle=\prod_{i=1}^L(S_i^\dag)^{N_i}|\theta\rangle \label{hierarchy:permanent:slater}
\end{equation} 
which are defined by means of the integer partitioning $[N_1,N_2,\dots,N_L]$ of $N$ pairs over $L$ spin-orbitals. Occupied orbitals are denoted by $N_i=1$, whereas empty orbitals have $N_i=0$.  The set of all possible states (\ref{hierarchy:permanent:slater}) spans a basis for the seniority-zero space of DOCI, so also the APIG state can be expanded in this basis
\begin{equation}
|\textrm{APIG}_N\rangle=\sum_{[N_1,\dots,N_L]}c^N_{[N_1,\dots,N_L]}|[N_1,N_2,\dots,N_L]\rangle
\end{equation}
It is fairly straightforward to show that the expansion coefficient $c^N_{[N_1,\dots,N_L]}$, or equivalently, the overlap of the APIG state with the seniority-zero Slater determinants $|[N_1,\dots,N_L]\rangle$ can be obtained as a \emph{permanent} \cite{muir:1960}
\begin{equation}
\langle[N_1,\dots,N_L]|\textrm{APIG}_N\rangle=\textrm{per}(G_{[N_1,\dots,N_L]})\label{hierarchy:permanent:overlap}
\end{equation} 
in which the notation $G_{[N_1,\dots,N_L]})$ is used for the reduced geminal matrix obtained by selecting only the columns $i$ for which $N_i=1$ from the geminal matrix $G$.  
\begin{equation}
G_{[N_1,\dots,N_L]}=\left(\begin{array}{cccc} G_{1i_1|N_{i_1}=1} & G_{1i_2|N_{i_2}=1} & \dots & G_{1i_N|N_{i_N}=1} \\  G_{2i_1|N_{i_1}=1} & G_{2i_2|N_{i_2}=1} & \dots & G_{2i_N|N_{i_N}=1}  \\ \vdots & \vdots & & \vdots \\  G_{Ni_1|N_{i_1}=1} & G_{Ni_2|N_{i_2}=1} & \dots & G_{Ni_N|N_{i_N}=1} \end{array}\right)\label{hierarchy:permanent:reducedgeminalmatrix}
\end{equation}
For instance, the overlap of the APIG state with the restricted Hartree-Fock state is given by the permanent of the reduced geminal matrix obtained by taking the first $N$ (occupied) columns from the geminal matrix
\begin{align}
\langle[\underbrace{1,\dots,1}_N,\underbrace{0,\dots,0}_{L-N}]|\textrm{APIG}_N\rangle&=\textrm{per}(G_{[\underbrace{1,\dots,1}_N,\underbrace{0,\dots,0}_{L-N}]}),\\
&=\textrm{per}\left(\begin{array}{cccc} G_{11} & G_{12} & \dots & G_{1N} \\  G_{21} & G_{22} & \dots & G_{2N}  \\ \vdots & \vdots & & \vdots \\  G_{N1} & G_{N2} & \dots & G_{NN} \end{array}\right).
\end{align}
Similar to the determinant of a matrix, the permanent of a square matrix is defined as the Laplace expansion over all possible permutations $\sigma\in S_N$ of $N$ elements
\begin{equation}
\textrm{per}(G) = \sum_{\sigma\in S_N}\left[ \prod_{i=1}^NG_{i\sigma(i)}\right],
\end{equation}  
with the main exception that the terms in the expansion are not weighted by the sign of the permutation $\sigma$.  This a significant feature of determinants, as it allows for the use of Gauss elimination to reduce its computation to polynomial scaling in all general cases.  Unfortunately for the permanent, there is no equivalent, leaving the Laplacian expansion the only option to compute the permanent of a general matrix, which comes with a factorial scaling as we must sum over all possible permutations in $S_N$ \cite{muir:1960}.  This puts geminal theory in a dire situation, as the computation of a simple overlap already invokes a factorially scaling computation.  Because of its tight relation with the permanent, geminal theory can be categorized in the \#P complexity class, one of the hardest identified complexity classes \cite{valiant:1979}.

The short proof of relation (\ref{hierarchy:permanent:overlap}) goes by induction.  It is trivially true for $N=1$
\begin{equation}
\langle [0,\dots,\underbrace{1}_{i},\dots 0] |\textrm{APIG}_1\rangle = \langle\theta| S_i\sum_{j=1}^L G_{1j}S_j^\dag|\theta\rangle = G_{1i}
\end{equation}
Let's assume (\ref{hierarchy:permanent:overlap}) holds for all APIG states with $N-1$ pairs.  Adding one extra generalized pair leads to
\begin{equation}
|\textrm{APIG}_N\rangle = \sum_{i=1}^N G_{N i}S_i^\dag|\textrm{APIG}_{N-1}\rangle
\end{equation}
Collecting all coefficients of the basis state $|[N_1,N_2,\dots,N_L]\rangle$ leads to the recursion relation
\begin{equation}
c^N_{[N_1,N_2,\dots,N_L]}=\sum_{i=1|N_i=1}^L G_{Ni}\times c^{N-1}_{[N_1,\dots,N_i-1,\dots,N_L]},
\end{equation}
in which the summation runs only over those orbitals that are occupied in $|[N_1,N_2,\dots,N_L]\rangle$.  Given that each $c^{N-1}_{[N_1,\dots,N_i-1,\dots,N_L]}$ is a permanent itself of a reduced geminal matrix (\ref{hierarchy:permanent:reducedgeminalmatrix}) for $N-1$ pairs with the $i$th column left out, this recursion relation is identical to the recursion relation for the permanent of the matrix for $N$ pairs.  This concludes the proof. 

\subsection{Variational approaches}
The factorial computational scaling of the permanent limits the application of geminal theory for larger systems, despite its intuitive mean-field character.  However, if the system can be partitioned into smaller tractable fractions, it is still possible to compute properties of the full system in a reasonable time.  This is the philosophy of the Antisymmetrized Product of Strongly-orthogonal Geminals (APSG), in which the full set of spin-orbitals system is subdivided into smaller subsets, each one with its own individual, and orthogonal, APIG wavefunctions \cite{surjan:1999}.  However more desirable from a computational point of view, imposing separability of the electron pairs introduces an additional bias into the preferred pairing structure that requires sufficient chemical insight from the practitioner.   This constraint can either be relaxed, for instance in the singlet-type strongly orthogonal \cite{cagg:2014} and $p$-orthogonal geminals \cite{cassamchenai:2023}, or taken into the extreme in the Generalized Valence Bond - Perfect Pairing (GVB-PP), in which each geminal is restricted to a resonance between a bonding-antibonding pair of molecular orbitals \cite{goddard:1973}. 

In the last decade, an alternative variational geminal theory has emerged on the premises of Richardson-Gaudin (RG) integrable model Hamiltonians \cite{johnson:2013,johnson:2020}, in which the geminals take the following form
\begin{equation}
|\textrm{RG}\rangle=\prod_{\alpha=1}^N\left[\sum_{i=1}^L\frac{S_i^\dag}{\varepsilon_i-x_\alpha}\right]|\theta\rangle.
\end{equation}
with the geminal matrix constrained to a Cauchy form in terms of two sets of parameters $\varepsilon_i$ and $x_\alpha$
\begin{equation}
G_{\alpha i }=\frac{1}{\varepsilon_i - x_\alpha}.
\end{equation}
The Cauchy form of the geminal matrix allows for the efficient evaluation of the permanent as a ratio of two determinants via Borchardt's theorem \cite{borchardt:1855,chavez:2024}, however this is not sufficient for a full-fledged variational theory.  For this to be the case, the two sets of parameters $\varepsilon_i$ and $x_\alpha$ need to be linked together by means of the RG equations \cite{dukelsky:2004a,ortiz:2005}.  Again, the main idea is keep the computation of the permanent in check, this time by tying it to the integrability of the underlying model Hamiltonian \cite{faribault:2008, claeys:2017a,moisset:2022}.  An interesting observation is that the Antisymmetrized Geminal Power (AGP) wavefunction \cite{coleman:1965}
\begin{equation}
|\textrm{APG}\rangle = \left[\sum_{i=1}^L G_iS_i^\dag\right]^N|\theta\rangle,
\end{equation}
or, equivalently, the number-projected Bardeen-Cooper-Schrieffer (BCS) state, can also be rephrased as an eigenstate of an integrable model, in particular the Moore-Read state of a factorizable RG model \cite{rombouts:2010}.  This observation underlines the richness and versatility of the AGP state for strongly correlated systems \cite{dutta:2020}, which happens to be the only geminal implemented on a quantum device to date \cite{khamoshi:2023}.
\subsection{The projected Schr\"odinger Equation}
As illustrated in the previous section, variational geminal theories are few and far between.  The feasibility of a tractable variational theory hinges on the tractability of the permanents involved in the computation.  For this, it is reasonable to explore whether the variational requirement can be relaxed in favour of ending up with a geminal formalism for which all involved permanents are tractable.  One approach is to solve the Schr\"odinger \emph{projectively} rather than variationally, as it typically only involves the computation of overlaps with respect to a selected set of Slater determinants, in this case seniority-zero Slater determinants.  The use of projective approaches has been popularized by Coupled Cluster theory \cite{bartlett:2007}, and was considered first for paired wavefunctions in the context of orbital optimized GVB-PP \cite{cullen:1996}.

The crux of the projected Schr\"odinger Equation (pSE) approach is to project the time-independent Schr\"odinger equation on at least as many reference states as there are free parameters in the wavefunction ansatz, and solve this subset of (non-linear) equations.  In the case of an over determined set of equations, the solution can be found by minimizing the mean square error with respect to the free parameters.  This approach is equivalent to the variance minimization procedure \cite{bartlett:1935} in the limit of a complete set of projection states.   

In the pSE approach, we require the Schr\"odinger equation 
\begin{equation}
 H|\Psi\rangle=E|\Psi\rangle,
\end{equation}
to be satisfied only with respect to a subset of reference states.   As reference states, we will first choose the state \cite{cullen:1996}
\begin{equation}
|\Phi_0\rangle=\prod_{j=1}^NS_j^\dag|\theta\rangle,\label{hierarchy:pse:phi0}
\end{equation}
as a base reference, which is a simple filling of the first $N$ spin-orbitals $j$ with $N$ pairs.  This state can be the restricted Hartree-Fock state, however any seniority-zero Slater determinant is viable, as one can always reorder any set of orbitals such that $|\Phi_0\rangle$ can be written as (\ref{hierarchy:pse:phi0}).   Second, one can now consider 1-pair excitations on top of $|\Phi_0\rangle$ as additional reference projection states 
\begin{equation}
|\Phi_i^a\rangle=S^\dag_a\prod_{j=1,(\neq i)}^NS_j^\dag |\theta\rangle,\label{hierarchy:pse:phiia}
\end{equation}
which corresponds to the promotion of a pair from an occupied level $i$ to an empty level $a$.  We will employ the notation $i,j,k,\dots$ for occupied levels, $a,b,c,\dots$ for virtual orbitals, and $p,q,r,\dots$ for general orbitals.    Using these reference states, we obtain
\begin{align}
&\langle\Phi_0|H|\Psi\rangle=E\langle\Phi_0|\Psi\rangle\label{hierarchy:pse:Edef1}\\
&\langle\Phi_i^a|H|\Psi\rangle=E\langle\Phi_i^a|\Psi\rangle\label{hierarchy:pse:Edef2}.
\end{align}
Eliminating $E$ from these equations gives rise to
\begin{equation}
\langle\Phi_0|\Psi\rangle \langle\Phi_i^a|H|\Psi\rangle=\langle\Phi_0|H|\Psi\rangle\langle\Phi_i^a|\Psi\rangle.\label{hierarchy:pse:formaleq} 
\end{equation}
At this point, no assumptions have been made on $|\Psi\rangle$, other than that it is a seniority-zero state.  To keep the discussion as general as possible at this point, we act with the Hamiltonian to the left on the reference states, rather than to the right on the yet unknown target state.  Using the DOCI Hamiltonian (\ref{seniority:doci:hamiltonian}) on the bra reference states $\langle\Phi_0|$ and $\langle\Phi_i^a|$ yields

\begin{align}
\langle\Phi_0|H&=\Big[\sum_{j=1}^N(2\varepsilon_j+V_{jj})+4\sum_{jk=1}^NW_{jk} \Big]\langle\Phi_0| + \sum_{j=1}^N\sum_{b=N+1}^LV_{jb}\langle\Phi_j^b|\label{pse:phi0H}\\
\langle\Phi^a_i|H&=\Big[\sum_{j=1}^N(2\varepsilon_j+V_{jj})+4\sum_{jk=1}^NW_{jk}+(2\varepsilon_a-2\varepsilon_i) + (V_{aa}-V_{ii})\label{pse:phiiaH}\\
&+4\sum_{j=1}^N(W_{ja}-W_{ji}+W_{aj}-W_{ij}) +4(W_{aa}-W_{ia}-W_{ai}+W_{ii})\Big]\langle\Phi_i^a|\notag\\
& +V_{ai}\langle\Phi_0|+\sum_{j=1,\neq i}^NV_{ji}\langle\Phi_j^a|+\sum_{b=N+1,\neq a}^LV_{ab}\langle\Phi_i^b|+\sum_{b=N+1,\neq a}^L\sum_{j=1,\neq i}^NV_{jb}\langle\Phi_{ij}^{ab}|,\notag
\end{align}
with $|\Phi^{ab}_{ij}\rangle$ a 2-pair excitation state.  Similarly, the action of the Hamiltonian on a 2-pair excitation will yield 3-pair excitations $\langle\Phi_{ijk}^{abc}|$, etc.\ which can all be systematically, however tediously, derived.  To emphasize the structure of the interactions, we have not yet used the symmetric properties of the $W=W^T$ interaction matrix, but will do so in the following.

Substituting these relations into (\ref{hierarchy:pse:formaleq}) gives rise to the set of equations
\begin{align}\label{hierarchy:pse:expleq}
&\Big[(2\varepsilon_a-2\varepsilon_i)+(V_{aa}-V_{ii})+8\sum_{j=1}^N(W_{aj}-W_{ij})\\
&\quad+4(W_{aa}-2W_{ia}+W_{ii})\Big]\langle\Phi_0|\Psi\rangle\langle\Phi_i^a|\Psi\rangle+V_{ia}[\langle\Phi_0|\Psi\rangle^2-\langle\Phi_i^a|\Psi\rangle^2]\\
&\quad+\sum_{j=1,\neq i}^N[V_{ji}\langle\Phi_0|\Psi\rangle-V_{ja}\langle\Phi^a_i|\Psi\rangle]\langle\Phi_j^a|\Psi\rangle\\
&\quad+\sum_{b=N+1,\neq a}^m[V_{ab}\langle\Phi_0|\Psi\rangle-V_{ib}\langle\Phi^a_i|\Psi\rangle]\langle\Phi^b_i|\Psi\rangle\\
&\quad+\sum_{b=N+1,\neq a}^m\sum_{j=1,\neq i}^NV_{jb}[\langle\Phi_0|\Psi\rangle\langle\Phi_{ij}^{ab}|\Psi\rangle-\langle\Phi_i^a|\Psi\rangle\langle\Phi_j^b|\Psi\rangle]=0
\end{align}
It is interesting to note that all free variables are now encoded into the overlaps of the target state with the reference projection states, underlining again the significance of the scaling of the permanent expression in geminal theory.  This observation is extended to the evaluation of the energy, as expressed in relation (\ref{hierarchy:pse:Edef1})
\begin{equation}\label{hierarchy:pse:energy}
E=\frac{\langle\Phi_0|H|\Psi\rangle}{\langle\Phi_0|\Psi\rangle}=\sum_{j=1}^{N}(2\varepsilon_j + V_{jj}) +4\sum_{jk=1}^NW_{jk} + \sum_{j=1}^N \sum_{b=N+1}^mV_{jb}\frac{\langle\Phi_j^b|\Psi\rangle}{\langle\Phi_0|\Psi\rangle}
\end{equation}
Although not immediately obvious, this relation should be equivalent with expression (\ref{hierarchy:pse:Edef2}) due to the formal consistency with relation (\ref{hierarchy:pse:formaleq}) of the pSE 
\begin{equation}\label{hierarchy:pse:energy:ia}
E=\frac{\langle\Phi_i^a|H|\Psi\rangle}{\langle\Phi_i^a|\Psi\rangle}.
\end{equation}
However, the energy expression (\ref{hierarchy:pse:energy:ia}) appears to be explicitly dependent on the pair-hole excitation $(i,a)$,  as 
\begin{align}
E&=\sum_{j=1}^{N}(2\varepsilon_j + V_{jj}) +4\sum_{jk=1}^NW_{jk}
+(2\varepsilon_a-2\varepsilon_i) +(V_{aa}-V_{ii})\notag\\
&+8\sum_{j=1}^N(W_{aj}-W_{ij})+4(W_{aa}-2W_{ia}+W_{ii})+V_{ai}\frac{\langle\Phi_0|\Psi\rangle}{\langle\Phi_i^a|\Psi\rangle}\\
&+\sum_{j=1\neq i}^NV_{ji}\frac{\langle\Phi_j^a|\Psi\rangle}{\langle\Phi_i^a|\Psi\rangle}+\sum_{b=N+1,\neq a }^LV_{ab}\frac{\langle\Phi_i^b|\Psi\rangle}{\langle\Phi_i^a|\Psi\rangle}+\sum_{j=1,\neq i}^N\sum_{b=N+1,\neq a}^LV_{jb}\frac{\langle\Phi_{ij}^{ab}|\Psi\rangle}{\langle\Phi_i^a|\Psi\rangle}.\notag
\end{align}
Only when phrased within the bi-variational principle (see section \ref{subsection:bivariational}) will it be unambiguously proven that the energy expressions (\ref{hierarchy:pse:energy}) and (\ref{hierarchy:pse:energy:ia}) are equivalent for all $(i,a)$ pair excitations. 
\subsection{Antisymmetrized Product of 1-reference orbital Geminals}
The central role of the overlap of the target state $|\Psi\rangle$ with the projection reference states $\{|\Phi_0\rangle,|\Phi_i^a\rangle\}$ is clear from relations (\ref{hierarchy:pse:expleq}) and (\ref{hierarchy:pse:energy}).  Therefore, it is key for a tractable geminal theory to propose ans\"atze that have a tractable permanent overlap with the reference state.  The Antisymmetrized Product of 1-reference orbital Geminals (AP1roG) \cite{limacher:2013,johnson:2013,boguslawski:2014a} was designed explicitly with this feature in mind
\begin{equation}\label{hierarchy:ap1rog:state}
|\textrm{AP1roG}\rangle = \prod_{\alpha=1}^N\left[S_\alpha^\dag + \sum_{i=N+1}^Lc_\alpha^i S_i^\dag\right]|\theta\rangle.
\end{equation}
Although envisioned initially within a hierarchy of $n$-reference orbitals \cite{limacher:2013}, the first $n=1$ rung on the ladder (AP1roG) turned out already to be remarkably effective for several reasons.  First, the computational scaling of the AP1roG wavefunction within the pSE approach is quartic in the size of the system $\mathcal{O}(N^2(L-N)^2)$, comparable to mean-field like methods.  Second, AP1roG is shown to be equivalent to paired Coupled Cluster Doubles (pCCD) \cite{stein:2014,henderson:2014b}. Finally third, extensive experimentation with the energy predictions from AP1roG for quantum chemistry applications showed that AP1roG energies are effectively indistinguishable from the exact energies from DOCI Hamiltonians \cite{tecmer:2014,shepherd:2016}.  

The essential structure behind the AP1roG wavefunction (\ref{hierarchy:ap1rog:state}) is that it fixes a set of orbitals as the reference seniority-zero Slater state, and leaves the virtual space free, allowing for an interpretation as a \emph{partially} orthogonal geminal \cite{neuscamman:2013}.  This is best illustrated by means of the geminal matrix 
\begin{equation}\label{ap1rog:geminal}
G=\left(\begin{array}{cccc:cccc}
       1 & 0 & \dots & 0 & c_1^{N+1} & c_1^{N+2} & \dots & c_1^L\\
       0 & 1 & \dots & 0 & c_2^{N+1} & c_2^{N+2} & \dots & c_2^L\\
       \vdots & \vdots & & \vdots & \vdots & \vdots & & \vdots \\
       0 & 0 & \dots & 1 & c_N^{N+1} & c_N^{N+2} & \dots & c_N^L
      \end{array}
\right).
\end{equation}
Although there is an undeniable bias associated with this choice, results show that the orbital optimization \cite{boguslawski:2014b,boguslawski:2014c} converges to pairs of bonding and antibonding pairs in the occupied/virtual sector of the geminal matrix, much alike GVB-PP wavefunctions. 

With this specific geminal, we obtain the following overlaps 
\begin{align}
 &\langle\Phi_0|\Psi\rangle=1,\\
 &\langle\Phi_i^a|\Psi\rangle=c_i^a,\\
 &\langle\Phi^{ab}_{ij}|\Psi\rangle=c_i^a c_j^b + c_i^b c_j^a,
\end{align}
which are quadratic expressions at worst, leading to the pSE equations {\ref{hierarchy:pse:expleq})
\begin{align}\label{hierarchy:ap1rog:pse}
&\Big[(2\varepsilon_a-2\varepsilon_i)+(V_{aa}-V_{ii})+8\sum_{j=1}^N(W_{aj}-W_{ij})+4(W_{aa}-2W_{ai}+W_{ii})\Big]c_i^a\notag\\
&\quad+V_{ia}[1-(c_i^a)^2]+\sum_{j=1,\neq i}^N[V_{ji}-V_{ja}c_i^a]c_j^a+\sum_{b=N+1,\neq a}^L[V_{ab}-V_{ib}c^a_i]c^b_i\notag\\
&\quad+\sum_{b=N+1,\neq a}^L\sum_{j=1,\neq i}^NV_{jb}c_i^bc_j^a=0.
\end{align}
Furthermore, the energy expression (\ref{hierarchy:pse:energy}) becomes a linear function in the geminal matrix
\begin{equation}
E=\sum_{j=1}^{N}(2\varepsilon_j + V_{jj}) +4\sum_{jk=1}^NW_{jk}+ \sum_{j=1}^N \sum_{b=N+1}^LV_{jb}c_j^b\label{hierarchy:ap1rog:energy}
\end{equation}
The set of pSE (\ref{hierarchy:ap1rog:pse}) for AP1roG is non-linear, necessitating the use of non-linear solvers, such as Newton-Raphson.  However, for weak interactions ($V_{jb}\ll (\varepsilon_a-\varepsilon_i)$ and $W_{jb}\ll (\varepsilon_a-\varepsilon_i)$ $\forall a,i,b,j$, one can assume that the reference state $|\Phi_0\rangle$ will be dominant and that all coefficients $c_i^a$ will be small.  By dividing the pSE (eq.\ \ref{hierarchy:ap1rog:pse}) by $(2\varepsilon_a-2\varepsilon_i)$ and comparing order by order, we obtain 
\begin{equation}\label{hierarchy:ap1rog:weak}
 c_{i}^a\approx-\frac{V_{ia}}{(2\varepsilon_a-2\varepsilon_i)} + \mathcal{O}(2)
\end{equation}
From this relation, we can infer that each reference spin orbital $i$ will resonate best with a virtual orbital $a$ for which either the pair scattering scattering matrix element is largest, or a quasi-degenerate orbital.  In addition, the relative minus sign strengthens the interpretation of the $(i,a)$ pair as resonating valence bond pair. 

As a conclusion, the AP1roG geminals in the pSE approach provide a fairly compact and elegant means to practice geminal theory at a reasonable cost, and we refer to Tecmer et.\ al.\ \cite{tecmer:2022} for a recent review on interesting applications. 

\section{Reduced density matrices}

$k$-Body reduced density matrices ($k$RDM) offer a compact way to represent correlated quantum systems, and allow for the computation of many experimental observables and quantum chemical quantities other than the energy, such as electric dipole transition amplitudes and bond orders respectively.  Furthermore, as the quantum chemical Hamiltonian is a two-body operator, knowledge of the 2RDM allows for the exact computation of the energy at a reduced computational cost \cite{hussimi:1940,lowdin:1955}.  To fix ideas, we will focus on the 2RDM, however extensions to other $k$RDMS are possible.  

In the seniority-zero framework, the only non-negative components of the 2RDM are given by
\begin{align}
\Pi_{pq}&=\langle\Psi|S_p^\dag S_q|\Psi\rangle\\
D_{pq}&=\langle\Psi|n_p n_q|\Psi\rangle,
\end{align}
in which the indices $p$ and $q$ run over all occupied and virtual orbitals.  Both matrices carry different interpretations as pair-scattering amplitudes ($\Pi$) and pair-correlation indices ($D$), however they are related to each other on the diagonal $p=q$, for which the following relation hold
\begin{equation}
D_{pp}= 2(N+\Pi_{pp}),
\end{equation} 

The 2DMs for most APIG states, including the AP1roG, inherit the computational intractability of general seniority-zero states.  This is because it involves the overlap of the geminal states with all possible seniority-zero Slater determinants in the Hilbert space.  Only the RG states escape this fate, however at the cost of associating the RG states with a so-called on-shell eigenstate of an integrable model \cite{faribault:2008, claeys:2017a,moisset:2022}.  For all other APIG state, one needs to resort to other methods, for instance projection methods.  However, 2RDMs are not unambiguously defined in a pSE approach without additional assumptions on the projected reference states.  In order to appreciate this, it is instructive to start with a ``poor man's version'' of the 2DM.   

\subsection{Poor man's 2RDM}

In analogy with the energy expression (\ref{hierarchy:pse:energy}), one can define the 2DMs as
\begin{align}
\Pi_{pq}=\frac{\langle \Phi_0|S_p^\dag S_q|\Psi\rangle}{\langle \Phi_0|\Psi\rangle},\quad D_{pq}=\frac{\langle \Phi_0|n_pn_q|\Psi\rangle}{\langle \Phi_0|\Psi\rangle},
\end{align}
leading to the (block) matrices
\begin{equation}
\Pi=\left(\begin{array}{c:c} \delta_{ij} & \frac{\langle\Phi_i^b|\Psi\rangle}{\langle\Phi_0|\Psi\rangle} \\ \hdashline 0 & 0 \end{array}\right)\quad D=\left(\begin{array}{c:c} 4 & 0 \\ \hdashline 0 & 0 \end{array}\right)\label{2RDM:poormanPD}
\end{equation}
which needs to be read as follows in occ/virt notation
\begin{align}
&\Pi_{ij}=\delta_{ij},\quad \Pi_{ib}=\frac{\langle\Phi_i^b|\Psi\rangle}{\langle\Phi_0|\Psi\rangle}, \quad \Pi_{aj}=\Pi_{ab}=0\\
&D_{ij}=4,\quad D_{ib}=D_{aj}=D_{ab}=0.
\end{align}
However, in contrast to the energy expression (\ref{hierarchy:pse:energy}), this definition is \emph{not} independent of the choice of reference state.  This is best illustrated with a simple example 
\begin{equation}
\tilde{\Pi}_{aj}=\frac{\langle \Phi_i^b|S_a^\dag S_j|\Psi\rangle}{\langle \Phi_i^b|\Psi\rangle}=\delta_{ij}\delta_{ab}\frac{\langle\Phi_0|\Psi\rangle}{\langle\Phi_i^b|\Psi\rangle}\neq\Pi_{aj}
\end{equation}
The reason for this ambiguity is that consistency of observables in the pSE is only guaranteed for the Hamiltonian, which was done by construction.  Therefore, it makes sense to start from a projection-state independent quantity, and derive the 2DM from that particular quantity.  If we choose the energy as that projection-independent quantity, we end up with Hellmann-Feynman theorem.

Unfortunately, the Hellmann-Feynman theorem is not simply valid for the pSE, because it is not variational.  Let $\alpha$ be a parameter in the Hamiltonian with an associated operator $\frac{\partial\hat{H}}{\partial\alpha}$, then
\begin{equation}
\frac{\partial E}{\partial\alpha}=\frac{\partial}{\partial \alpha}\frac{\langle\Phi_0|H|\Psi\rangle}{\langle\Phi_0|\Psi\rangle}=\frac{\langle\Phi_0|\frac{\partial H}{\partial\alpha}|\Psi\rangle}{\langle\Phi_0|\Psi\rangle}+\frac{\langle\Phi_0|H-E|\frac{\partial\Psi}{\partial\alpha}\rangle}{\langle\Phi_0|\Psi\rangle}\label{2DM:pse:hellmannfeynman}
\end{equation}
Opposed to the conventional Hellmann-Feynman theorem, there is no reason why the last term should magically vanish.   The reason is that $|\Psi\rangle$ has not been obtained variationally, which is the prerequisite for the Hellmann-Feynman theorem to be valid.  So, we want a theoretical framework that is both variational (to ensure the Hellmann-Feynman theorem) and skew (to allow for the pSE) at the same time.  The natural answer to this is the \emph{bi}-variational principle, first introduced in quantum chemistry by Boys and Handy \cite{boys:1969}.  It has been used explicitly in geminal theory in the context of off-shell RG geminals \cite{johnson:2022} and implicitly for AP1roG/pCCD for orbital optimization purposes \cite{boguslawski:2014c,henderson:2014b}.

\subsection{The bi-variational principle }\label{subsection:bivariational}

The \emph{bi}-variational framework is a generalization of the variational principle, because it starts from a \emph{bi}-functional expression of the energy
\begin{equation}
E[\eta,\theta]=\frac{\langle\Phi[\eta]|H|\Psi[\theta]\rangle}{\langle\Phi[\eta]|\Psi[\theta]\rangle},
\end{equation}
dependent on both a left state $\langle\Phi[\eta]|$ and right state $|\Psi[\theta]\rangle$.  Variation of the energy bi-functional to the left and right states leads to 
\begin{align}
\tfrac{\partial E}{\partial \theta}=0&\rightarrow \langle\Phi|H-E|\tfrac{\partial\Psi}{\partial\theta}\rangle=0\\
\tfrac{\partial E}{\partial \eta}=0&\rightarrow \langle\tfrac{\partial\Phi}{\partial\eta}|H-E|\Psi\rangle=0
\end{align}
A beneficial feature of the bi-variational principle is that  whenever the exact eigenstates are on the manifolds $\langle\Phi[\eta]|$ or $|\Psi[\theta]\rangle$, they emerge as a solution.

Because of the variational character of the energy, the Hellmann-Feynman theorem is again valid
\begin{equation}
\frac{\partial E}{\partial\alpha}=\frac{\langle\Phi|\frac{\partial H}{\partial\alpha}|\Psi\rangle}{\langle\Phi|\Psi\rangle}+\frac{\partial E}{\partial\eta}\frac{\partial\eta}{\partial\alpha}+\frac{\partial E}{\partial\theta}\frac{\partial\theta}{\partial\alpha}=\frac{\langle\Phi|\frac{\partial H}{\partial\alpha}|\Psi\rangle}{\langle\Phi|\Psi\rangle}\label{2DM:bivar:hellmannfeynman}
\end{equation}
This means that the calculation of the observables $\frac{\partial H}{\partial \alpha}$ is now unambiguously defined via the variationally obtained $\langle\Phi[\eta]|$ and $|\Psi[\theta]\rangle$.  The question is now if the pSE is consistent with the bi-variational principle

\subsubsection{the pSE from the bi-variational principle}

It is straightforward to show that the pSE is consistent with the bi-variational approach, whenever a linear expansion in the projection states is chosen for the left state 
\begin{equation}
\langle\Phi[\eta]|=\eta_0\langle\Phi_0|+\sum_{i=1}^N\sum_{a=N+1}^L\eta_{i}^a\langle \Phi_i^a|,
\end{equation}
Leaving the specific structure of $|\Psi\rangle$ unspecified at this moment, left-variation of the functional leads to 
\begin{align}
\tfrac{\partial E}{\partial \eta_0}=0&\rightarrow \langle\Phi_0|H-E|\Psi\rangle=0,\label{2DM:bivar:lefteq1}\\
\tfrac{\partial E}{\partial \eta_i^a}=0&\rightarrow \langle\Phi_i^a|H-E|\Psi\rangle=0,\label{2DM:bivar:lefteq2}
\end{align}
which is exactly the set of pSE equations (\ref{hierarchy:pse:Edef1}-\ref{hierarchy:pse:Edef2}).  Because of the linearity of the left trial state $\langle\Phi[\eta]|$, the variables $\eta$ do not occur in the pSE equations.  To determine these coefficients, one must also consider the right variation.  For this, it is necessary to allocate a functional behaviour to the target state $|\Psi[\theta]\rangle$.  To fix ideas, we continue with AP1roG, in which the geminal matrix elements $c_i^a$ now function as \emph{variational} parameters
\begin{equation}
|\Psi\rangle=\prod_{i=1}^N\left(S_i^\dag +\sum_{a=N+1}^Lc_i^aS_a^\dag\right)|\theta\rangle.
\end{equation}
The right variation equations are given by
\begin{equation}
\tfrac{\partial E}{\partial c_i^a}=0\rightarrow \langle\Phi|H-E|\tfrac{\partial\Psi}{\partial c_i^a}.\rangle=0\label{2DM:bivar:righteq}
\end{equation}
It is remarkable that the derivatives of the geminal wrt the coeficients again yields another geminals
\begin{equation}
|\tfrac{\partial \Psi}{\partial c_i^a}\rangle=\prod_{j=1,\neq i}^N\left(S_j^\dag+\sum_{b=N+1,\neq a}^Lc_j^bS_b^\dag\right)S_a^\dag|\theta\rangle.
\end{equation}
However, this geminal is not strictly an AP1roG, because the reference orbital $i$ is not present  The overlaps with the projection states are given by
\begin{align}
\langle\Phi_0|\tfrac{\partial \Psi}{\partial c_i^a}\rangle=&0,\\
\langle\Phi_j^b|\tfrac{\partial \Psi}{\partial c_i^a}\rangle=&\delta_{ij}\delta_{ab},\\
\langle\Phi_{jk}^{bd}|\tfrac{\partial \Psi}{\partial c_i^a}\rangle=&[\delta_{ij}\delta_{ab}c_k^d+\delta_{ik}\delta_{ad}c_j^b+\delta_{ij}\delta_{ad}c_k^b+\delta_{ik}\delta_{ab}c_j^d](1-\delta_{bd})(1-\delta_{jk}),\notag
\end{align}
in which the last factors correct for the fact that one cannot excite and occupy a level twice ($i\neq k$, $b\neq d$).  Together with the left-action relations of the Hamiltonian (\ref{pse:phi0H}-\ref{pse:phiiaH}), we get
\begin{align}
\langle\Phi_0&|H-E|\tfrac{\partial \Psi}{\partial c_i^a}\rangle=V_{ia},\\
\langle\Phi_j^b&|H-E|\tfrac{\partial \Psi}{\partial c_i^a}\rangle=(\tilde{\varepsilon}_a-\tilde{\varepsilon}_i)\delta_{ij}\delta_{ab}+V_{ij}\delta_{ab}+V_{ba}\delta_{ij}-\delta_{ij}\delta_{ab}(V_{ab}+V_{ij})\notag\\
&+(1-2\delta_{ab})(1-2\delta_{ij})V_{ia}c_j^b+(1-2\delta_{ab})\delta_{ij}\sum_{k=1}^NV_{ka}c_k^b\\
&+(1-2\delta_{ij})\delta_{ab}\sum_{d=N+1}^LV_{id}c_j^d\notag,
\end{align}
where we have introduced the short-hand notation
\begin{align}
(\tilde{\varepsilon}_a-\tilde{\varepsilon}_i)=&2(\varepsilon_a-\varepsilon_i)+(V_{aa}-V_{ii})+4(W_{aa}-W_{ia}-W_{ai}+W_{ii})\\
&+4\sum_{k=1}^N(W_{ak}-W_{ik}) + 4\sum_{k=1}^N(W_{ka}-W_{ki}).
\end{align}
Taking into account that 
\begin{equation}
\langle\Phi|H-E|\tfrac{\partial\Psi}{\partial c_i^a}\rangle=\Big[\eta_0\langle\Phi_0|+\sum_{jb}\eta_j^b\langle\Phi_j^b|\Big](H-E)|\tfrac{\partial\Psi}{\partial c_i^a}\rangle,
\end{equation}
the right-variational equations (\ref{2DM:bivar:righteq}) become
\begin{align}
&\Big[\eta_0+\sum_{jb}\eta_j^bc_j^b\Big]V_{ia}\notag\\
&+\eta_{i}^a\Big[(\tilde{\varepsilon}_i-\tilde{\varepsilon}_a)-(V_{aa}+V_{ii})+4V_{ia}c_i^a-2\sum_{k}V_{ka}c_k^a-2\sum_{b}V_{ib}c_i^b\Big]\notag\\
&+\sum_{j}\eta_j^a\Big[V_{ij}-2V_{ia}c_j^a+\sum_{b}V_{ib}c_j^b\Big]\notag\\
&+\sum_{b}\eta_i^b\Big[V_{ab}-2V_{ia}c_i^b+\sum_{k}V_{ka}c_k^b\Big]=0,\qquad \forall i,a.
\end{align}
Given that the $c$ coefficients have already been determined by the left-variational equations (\ref{2DM:bivar:lefteq1}-\ref{2DM:bivar:lefteq2}), this set of equations is linear in the $\{\eta_0,\eta_i^a\}$ coefficients.  Note however that we have $N(L-N)$ equations for $N(L-N)+1$ variables.  This extra degree of freedom can be eliminated by fixing the norm.  Two viable options are 
\begin{enumerate}
\item $\eta_0=1$.  This is equivalent with the variational OO approach for AP1roG described in \cite{boguslawski:2014c} with the left-variational parameters rephrased as Lagrange multipliers, or the pCCD approach described in \cite{henderson:2014b} where the left-variational state $\langle\Phi|$ is written in Coupled-Cluster language
\begin{equation}
\langle\Phi|=\langle\Phi_0|(1+Z),
\end{equation}
with $Z$ a $T_2$ like operator
\begin{equation}
Z=\sum_{i=1}^N\sum_{a=N+1}^Lz_i^aS^\dag_iS_a.
\end{equation}
The $z_i^a$ variables take the role of the present $\eta_i^a$ variables.
\item $\langle\Phi|\Psi\rangle=1$.  Here, the norm is fixed directly, leading to the relation
\begin{equation}
\langle\Phi|\Psi\rangle=\eta_0+\sum_{j=1}^N\sum_{b=N+1}^L\eta_j^bc_j^b=1.
\end{equation}
\end{enumerate}
\subsubsection{Observables in the bi-variational approach}
As soon as the $\eta$ parameters have been determined, we have a left state $\langle\Phi|$ which can be used to unambiguously calculate expectation values of operators, independent of the chosen projection state
\begin{equation}
\langle \hat{O}\rangle = \frac{\langle \Phi|\hat{O}|\Psi\rangle}{\langle\Phi|\Psi\rangle}
\end{equation}
It is worth noting that the pSE energy is not altered by reformulating the pSE in the bi-variational procedure, because of the persistence of eqs. (\ref{2DM:bivar:lefteq1}-\ref{2DM:bivar:lefteq2}).
\begin{equation}
\frac{\langle \Phi|H|\Psi\rangle}{\langle\Phi|\Psi\rangle}=\frac{\eta_0\langle \Phi_0|H|\Psi\rangle+\sum_{ia}\eta_i^a\langle\Phi_i^a|H|\Psi\rangle}{\langle\Phi|\Psi\rangle}=E.
\end{equation}
As a result, the energy $E$ remains independent of $\{\eta_0,\eta_i^a\}$, as is given by expression (\ref{hierarchy:pse:energy}).
\subsubsection{2RDMs in the bi-variational approach}
The 2RDMS can be viewed as a special case of operator.  To fix ideas, we will work in the norm constraint $\langle\Phi|\Psi\rangle=1$ convention, so
\begin{equation}
\Pi_{pq}=\langle\Phi|S_p^\dag S_q|\Psi\rangle,\quad D_{pq}=\langle\Phi|n_pn_q|\Psi\rangle
\end{equation}
We have to break down the calculation of $\Pi_{pq}$ into occupied/virtual parts, giving rise to (in order of increading complexity)
\begin{align}
\Pi_{ak}=&\eta_k^a,\quad \Pi_{ad}=\sum_{j}\eta_j^ac_j^d\\
\Pi_{id}=&\eta_0c_i^d+\sum_{j\neq i}\sum_{b\neq d}\eta_j^b(c_i^bc_j^d+c_i^dc_j^b)\\
\Pi_{ik}=&\delta_{ik}[1-2\sum_{b}\eta_k^bc_i^b]+\sum_{b}\eta_k^bc_i^b,
\end{align}
or in symbolic matrix form
\begin{equation}
\Pi=\left(\begin{array}{c:c} \delta_{ij}[1-2\sum_{b}\eta_k^bc_i^b]+\sum_{b}\eta_k^bc_i^b & \eta_0c_i^d+\sum_{j\neq i}\sum_{b\neq d}\eta_j^b(c_i^bc_j^d+c_i^dc_j^b) \\ \hdashline \eta_k^a & \sum_{j}\eta_j^ac_j^d \end{array}\right)\notag
\end{equation}
%5
Note that for $\{\eta_0,\eta_i^a\}=\{1,0\}$, $\Pi$ reduces back to the poor man's version (\ref{2RDM:poormanPD}).
Similarly, the $D_{pq}$ 2DM needs to be broken down into occupied/virtual parts.  They are given as follows
\begin{align}
D_{ad}&=4\delta_{ad}\sum_{j=1}^N\eta_j^ac_j^a,\quad D_{ak}=4\sum_{j\neq k}^N\eta_j^ac_j^a,\\
D_{ik}&=4[1-\sum_{b=N+1}^L(\eta_i^bc_i^b+\eta_k^bc_k^b-\delta_{ik}\eta_i^bc_k^b)],
\end{align}
or in matrix form
\begin{equation}
D=\left(\begin{array}{c:c} 4[1-\sum_{b}(\eta_i^bc_i^b+\eta_k^bc_k^b-\delta_{ik}\eta_i^bc_k^b)] & 4\sum_{j\neq i}^N\eta_j^dc_j^d \\ \hdashline 4\sum_{j\neq k}^N\eta_j^ac_j^a & 4\delta_{ad}\sum_{j=1}^N\eta_j^ac_j^a \end{array}\right).
\end{equation}
It is worth realizing that the 1 in the formula for $\Pi_{ik}$ and $D_{ik}$ are coming from the normalisation $1=\langle\Phi|\Psi\rangle=\eta_0+\sum_{jb}\eta_j^bc_j^b$.  A good consistency check is performed by calculating the energy from the 2DM.  It is a tedious but straightforward calculation to verify that
\begin{align}
E=&\sum_{p=1}^L\langle \Phi|n_j|\Psi\rangle + \sum_{pq=1}^LV_{pq}\langle\Phi|S_p^\dag S_q|\Psi\rangle + \sum_{pq=1}^LW_{pq}\langle\Phi|n_pn_q|\Psi\rangle,
\end{align}
is equivalent to expression (\ref{hierarchy:ap1rog:energy}).  Along the calculation, one needs to take into account that the $c$ coefficients solve the pSE equations (\ref{hierarchy:ap1rog:pse}).
\subsubsection{2RDMs via the Hellmann-Feynman theorem}
Ironically, the extra work calculating the $\{\eta_0,\eta_i^a\}$ parameters is not strictly \emph{necessary} now the bi-variational framework has been established.  Indeed, the Hellmann-Feynman theorem is now valid (\ref{2DM:bivar:hellmannfeynman}), so we can obtain the 2DMs from
\begin{equation}
\langle \Phi|S_p^\dag S_q|\Psi\rangle=\frac{\partial E}{\partial V_{pq}},\quad \langle \Phi|n_p n_q|\Psi\rangle=\frac{\partial E}{\partial W_{pq}}.
\end{equation}
The catch is now that the energy expression is only explicitly dependent on the $c_i^a$ coefficients, and \emph{not} the $\eta_i^a$ coefficients.  Breaking down the partial derivatives wrt occupied/virtuals, we obtain (immediately in matrix form)
\begin{equation}
\Pi=\left(\begin{array}{c:c} \delta_{ik}+\sum_{jb}V_{jb}\frac{\partial c_j^b}{\partial V_{ik}} & c_i^d+\sum_{jb}V_{jb}\frac{\partial c_j^b}{\partial V_{id}} \\ \hdashline \sum_{jb}V_{jb}\frac{\partial c_j^b}{\partial V_{ak}} & \sum_{jb}V_{jb}\frac{\partial c_j^b}{\partial V_{ad}} \end{array}\right).\label{2DM:bivar:hellmannfeynman:P}
\end{equation}
Note that $\Pi$ is not necessarily symmetric, and that it reduces (again) to the poor man's version (\ref{2RDM:poormanPD}) when the coefficients would be fixed constant.\newline
Similarly, we get for 
\begin{equation}
D=\left(\begin{array}{c:c} \delta_{ik}+\sum_{jb}V_{jb}\frac{\partial c_j^b}{\partial W_{ik}} & \sum_{jb}V_{jb}\frac{\partial c_j^b}{\partial W_{id}} \\ \hdashline \sum_{jb}V_{jb}\frac{\partial c_j^b}{\partial W_{ak}} & \sum_{jb}V_{jb}\frac{\partial c_j^b}{\partial W_{ad}} \end{array}\right).\label{2DM:bivar:hellmannfeynman:D}
\end{equation}
To get the derivatives $\frac{\partial c_j^b}{\partial V_{pq}}$, we only need to take the pSE equations, and derive them with respect to $V_{pq}$.  This will yield a linear set of equations in the variables $\frac{\partial c_j^b}{\partial V_{pq}}$ which can be solved easily numerically via matrix inversion.

Although the 2RDMS from the bivarational principle are significantly improved over the poor man's version, they still lack basic $N$-representability requirements, such as Hermiticity, due to the inherit skew-symmetric properties of the bi-variational principle.  It is not feasible to completely fix this deficiency, however algorithms exist \cite{lanssens:2018} to make the 2RDMs approximately $N$-representable by identifying the closest 2RDM that fulfills the so-called $\mathcal{PQG}$ conditions, however at the cost of a positive energy correction with respect to the original AP1roG energy (\ref{hierarchy:pse:energy})   

\section{Conclusions}\label{section:conclusions}

The present manuscript provides an overview of the mathematical structure behind the APIG and AP1roG/pCCD geminal wavefunctions which have enjoyed increased attention in the last decade.  The seniority framework is introduced as an optimal setting for geminal theory based on its Lie-algebraic features, and the hierarchy of geminal states is discussed within this framework.  The key role of the \emph{permanent} as the overlap between geminal states and seniority-zero Slater determinants is highlighted, and several approaches to avoid the pernicious factorial scaling have been discussed.  Much attention has been paid to the projected Schr\"odinger Equation approach, and how it can be derived consistently within the bi-variational principle.  The hope is that these notes can provide insight in the fundamental strengths and limitations of geminal theory and that it may inspire more breakthroughs towards a full-fledged APG geminal theory, tying in completely with the Lewis picture of chemical bonding without computational restrictions. 

\section*{Acknowledgements}

Over the past decade, the authors very much enjoyed conversations with Paul Ayers, Kasia Boguslawski, Patrick Bultinck, Pieter Claeys, Paul Johnson, Peter Limacher, Pawel Tecmer and Mario Van Raemdonck.  Funding from the FWO-Vlaanderen and Canada Research Chairs program is gratefully acknowledged.

%\thebibliography

\bibliographystyle{elsarticle-num}

%%
%\bibliography{bibstijn-012,
%              bibstijn-abc,
%              bibstijn-def,
%              bibstijn-gh,
%              bibstijn-ijk,
%              bibstijn-lmn,
%              bibstijn-opq,
%              bibstijn-rst,
%              bibstijn-uvw}

\end{document}